\documentclass[prd,twocolumn,eqsecnum,showpacs,amsfonts,amssymb]{revtex4}

\usepackage{graphicx}

\usepackage{bm}

\newcommand{\beq}{\begin{equation}}
\newcommand{\eeq}{\end{equation}}
\newcommand{\beqs}{\begin{eqnarray}}
\newcommand{\eeqs}{\end{eqnarray}}
\newcommand{\lsim}{\mathrel{\raisebox{-
.6ex}{$\stackrel{\textstyle<}{\sim}$}}}

\begin{document}

\title{Generalized Scheme Transformations for the Elimination of Higher-Loop
  Terms in the Beta Function of a Gauge Theory} 

\author{Robert Shrock}

\affiliation{C. N. Yang Institute for Theoretical Physics \\
Stony Brook University, Stony Brook, NY 11794 }

\begin{abstract}

We construct and study a generalized one-parameter class of scheme
transformations, denoted $S_{R,m,k_1}$ with $m \ge 2$, with the property that
an $S_{R,m,k_1}$ scheme transformation eliminates the $\ell$-loop terms in the
beta function of a gauge theory from loop order $\ell=3$ to order
$\ell=m+1$, inclusive. These scheme transformations are applied to the
higher-loop calculation of the infrared zero of the beta function of an
asymptotically free gauge theory with multiple fermions. We show that scheme
transformations in this generalized class satisfy a set of criteria for
physical acceptability over a larger range of numbers of fermions than
previously studied scheme transformations.  We also present an interesting
modification of a different type of scheme transformation that removes the
three-loop term in the beta function.

\end{abstract}


\maketitle


\section{Introduction}
\label{intro}

A basic property of a gauge theory is the dependence of the gauge coupling
$g=g(\mu)$ on the Euclidean momentum scale, $\mu$, where it is measured.  This
is described by the beta function of the theory, $\beta_g = dg/dt$ or
equivalently, $\beta_\alpha = d\alpha/dt = [g/(2\pi)]\beta_g$, where $dt=d\ln
\mu$ and $\alpha(\mu) = g(\mu)^2/(4\pi)$. The terms at loop order $\ell \ge 3$
in the beta function are dependent on the scheme used for regularization and
renormalization. Hence, one expects that, at least for sufficiently small
coupling, it is possible to carry out a scheme transformation that eliminates
these terms and yields a beta function with only one- and two-loop terms
\cite{thooft77}.  In \cite{sch} with T. Ryttov, we constructed and studied
explicit scheme transformations that remove terms at loop order $\ell \ge 3$
from the beta function.

An important application of such scheme transformations is to the analysis of
zero(s) of the beta function.  The beta function of an asymptotically free
non-Abelian gauge theory has an ultraviolet (UV) zero at $\alpha=0$, which is
an ultraviolet fixed point (UVFP) of the renormalization group (RG). If the
theory contains sufficiently many fermions, the (perturbatively calculated)
beta function may also have a infrared (IR) zero at a point $\alpha_{IR} >
0$. Depending how large $\alpha_{IR}$ is, this zero is either an exact or
approximate infrared fixed point (IRFP) of the renormalization group. Since the
terms of loop order $\ell \ge 3$ in the beta function are scheme-dependent, so
is the value of the IR zero when calculated to three-loop or higher-loop order.
In order to understand the physical implications of this IR zero, it is
necessary to assess the effect of scheme dependence on its value.  A study of
this dependence was carried out in \cite{sch} using several scheme
transformations.  In \cite{sch} we pointed out a set of criteria that a scheme
transformation must satisfy in order to be physically acceptable, and showed
that although it is straightforward for a scheme transformation to satisfy
these criteria in the vicinity of a zero of the beta function at $\alpha=0$,
they are a significant restriction on the choice of an acceptable scheme
transformation that can be applied at a generic infared zero of the beta
function.  Examples of scheme transformations were given in \cite{sch} that are
acceptable for small $\alpha$ but produce unphysical effects when applied at a
generic IR zero of the beta function.

One type of procedure that would be natural for a quantitative study of
scheme-dependence of a zero of the beta function would be to construct and
apply a scheme transformation that would remove successively higher and
higher-loop terms in the beta function and, at each stage, determine how this
removal shifted the position of the IR zero.  Extending the results of
\cite{sch}, in \cite{sch2} we defined a set of scheme transformations $S_{R,m}$
with $m \ge 2$ that remove the terms in the beta function at loop order
$\ell=3$ to $\ell=m+1$, inclusive and determined the range of $\alpha$ over
which $S_{R,2}$ and $S_{R,3}$ can be applied to study the IR zero of the beta
function of an asymptotically free gauge theory while satisfying the criteria
to avoid introducing unphysical pathologies.  For both $S_{R,2}$ and $S_{R,3}$
it was shown that these ranges are rather limited, which, in turn, restricts
one's ability to use these scheme transformations to study the
scheme-dependence of a zero of the beta function away from $\alpha=0$.

In this paper we present a generalized one-parameter class of scheme
transformation, denoted $S_{R,m,k_1}$ with $m \ge 2$, with the property that an
$S_{R,m,k_1}$ scheme transformation eliminates the $\ell$-loop terms in the
beta function of a quantum field theory from loop order $\ell=3$ to order
$\ell=m+1$, inclusive.  We give a detailed analysis of the application of this
scheme transformation to the infrared zero of an asymptotically free gauge
theory with gauge group $G={\rm SU}(N_c)$ and $N_f$ massless fermions in the
fundamental representation, and we show that it satisfies the physical
acceptability criteria specified in \cite{sch} over a wider range of $N_f$ and
hence a wider range of values of an infrared zero, $\alpha_{IR}$, than those
constructed and analyzed in \cite{sch}-\cite{sch2}.  We also investigate an
interesting modification of the $S_1$ scheme transformation presented in
\cite{sch}.

This paper is organized as follows. In Sect. \ref{basics} we recall some basic
information and notation that will be needed for our analysis.  In
Sect. \ref{srmk1} we define the scheme transformation $S_{R,m,k_1}$.  We
display explicit expressions for the resultant coefficients in the beta
function resulting from the application of the $S_{R,m,k_1}$ transformation in
Sect. \ref{bpsrmk1}. In Sects. \ref{sr2k1_application} and
\ref{sr3k1_application} we present specific results on the application of the
respective scheme transformations $S_{R,2,k_1}$ and $S_{R,3,k_1}$ to an IR zero
in the beta function of an SU($N_c$) gauge theory.  In Sect. \ref{lnn_section},
we give further results on the application of these scheme transformations in
the limit $N_c \to \infty$ and $N_f \to \infty$ with the ratio $N_f/N_c$ fixed.
In Sect. \ref{s1modified} we discuss a modification of a different type of
scheme transformation, namely the $S_1$ transformation of \cite{sch}.  We
present our conclusions in Sect. \ref{conclusions}.  Some additional results 
are included in appendices.


\section{Basics }
\label{basics} 

In this section we recall some basic formalism and notation that will be used
in our analysis.  The scheme transformation $S_{R,m,k_1}$ that we construct and
study can be applied to any gauge theory, vectorial or chiral, and non-Abelian
or Abelian.  Indeed, this transformation can also be applied to a quantum field
theory that does not involve gauge fields, with an appropriate replacement of
$g$ by the relevant interaction coupling.  Here we will focus on the
application to a vectorial non-Abelian gauge theory with gauge group $G$ and a
set of $N_f$ massless fermions transforming according to a representation $R$
of $G$.  Since these theories are vectorial, the gauge invariance would allow
nonzero fermion masses. However, in studying the evolution of the gauge
coupling as a function of the scale $\mu$, as this scale decreases below the
value of a given fermion mass, one would construct a low-energy effective field
theory by integrating this fermion out, so this massive fermion would not
affect the evolution of the coupling for scales below its mass. Hence, our
assumption of massless fermions does not entail a loss of generality.

It will be convenient to define the quantity
\beq
a(\mu) \equiv \frac{\alpha(\mu)}{4\pi} = \frac{g(\mu)^2}{16\pi^2} \ . 
\label{a}
\eeq
(The argument $\mu$ will often be suppressed in the notation.)  The function 
$\beta_\alpha$ function has the power-series expansion
\beq
\beta_\alpha = -2\alpha \sum_{\ell=1}^\infty b_\ell \, a^\ell =
-2\alpha \sum_{\ell=1}^\infty \bar b_\ell \, \alpha^\ell \ ,
\label{beta}
\eeq
where $\ell$ labels the loop order, $\bar b_\ell = b_\ell/(4\pi)^\ell$, and we
have extracted a minus sign so that the one-loop coefficient $b_1$ is positive
if the theory is asymptotically free. The $n$-loop ($n\ell$) $\beta$ function,
denoted $\beta_{\alpha,n\ell}$, is obtained from Eq. (\ref{beta}) by replacing
the upper limit on the $\ell$ loop summation by $n$ instead of $\infty$.  The
(scheme-independent) one-loop and two-loop coefficients $b_1$ and $b_2$ were
calculated in \cite{b1} and \cite{b2,jones}, respectively, and are listed for
reference in Appendix \ref{bell}.  As mentioned above, the $b_\ell$ with $\ell
\ge 3$ are scheme-dependent \cite{gross75,khuri}.  For a non-Abelian gauge
theory, $b_3$ and $b_4$ were calculated in \cite{b3} and \cite{b4} in the
modified minimal subtraction scheme \cite{msbar}. The property of asymptotic
freedom, i.e., $b_1 > 0$, requires that $N_f < N_{f,b1z}$, where
$N_{f,b1z}=11C_A/(4T_f)$ \ \cite{casimir}.  We assume that this condition is
satisfied.

If an asymptotically free gauge theory has sufficiently many massless fermions,
the beta function can exhibit an IR zero at a certain value, denoted
generically as $\alpha_{IR}$ \cite{b2,bz}.  As is evident from Eq. (\ref{b2}),
for small $N_f$, $b_2$ is positive, but it decreases with increasing $N_f$ and
passes through zero to negative values as $N_f$ increases through the value
\beq
N_{f,b2z} = \frac{34C_A^2}{4(5C_A+3C_f)T_f} \ .
\label{nfb2z}
\eeq
Since $N_{f,b2z} < N_{f,b1z}$, there is always an interval $I$, defined by 
\beq
I: \quad N_{f,b2z} < N_f < N_{f,b1z} \ , 
\label{nfinterval}
\eeq
in which the two-loop beta function, $\beta_{\alpha,2\ell}$, has an IR zero.
For $N_f \in I$, this zero of $\beta_{\alpha,2\ell}$ occurs at the
(scheme-independent) value
\beq
\alpha_{IR,2\ell} = 4\pi a_{IR,2\ell} = -\frac{4\pi b_1}{b_2} \ . 
\label{alfir_2loop}
\eeq
Henceforth, for definiteness, we focus on the case where the gauge group is
$G={\rm SU}(N_c)$ and the $N_f$ fermions transform according to the 
fundamental representation.

If the IR zero of the beta function occurs at a small value of the gauge
coupling, then this is an exact IR fixed point (IRFP) of the renormalization
group. With decreasing $N_f$, $\alpha_{IR}$ increases, eventually to a value at
which the gauge interaction is strong enough to trigger the formation of
bilinear fermion condensates with associated spontaneously chiral symmetry
breaking (S$\chi$SB). As a consequence of this, the fermions gain dynamical
masses of order the S$\chi$SB scale, denoted $\Lambda$. In the low-energy
effective field theory applicable at scales $\mu < \Lambda$, these fermions are
integrated out, the beta function changes to one with $N_f=0$, and the
resultant low-energy theory does not have an IR zero in its (perturbative) beta
function.  Thus, in this case, the initial zero is only an approximate, rather
than exact, fixed point of the renormalization group.  The value of
$N_f$ that separates these two regimes of infrared behavior is denoted
$N_{f,cr}$.  If the beta function of a theory has an IR zero that is only
slightly greater than the minimum value for fermion condensation, then the UV
to IR evolution exhibits slowly running, quasi-scale-invariant behavior over 
a substantial interval of scales $\mu$. This behavior, and the resultant
approximate Nambu-Goldstone boson (the dilaton) that results from the
spontaneous breaking of scale invariance by the bilinear fermion condensate,
might be relevant for physics beyond the Standard Model \cite{wtc}. 

Since $N_{f,cr}$ corresponds to a value $\alpha \sim O(1)$ for the exact or
approximate infrared zero of the beta function, one is motivated to calculate
this value to higher-loop order \cite{gkgg}. This was done in \cite{bvh,ps} for
this zero of the beta function and for the corresponding value of the anomalous
dimension of the fermion bilinear for a general gauge group and fermion
representation.  Additional higher-loop results on structural
properties of the beta function were calculated in \cite{bfs}-\cite{lnn}. 
In turn, this motivated the study of the scheme dependence of the IR zero in
beta in \cite{sch}-\cite{sch2} (some related work is in
\cite{dbeta}-\cite{lnf}.) 

A scheme transformation can be expressed as a mapping between $\alpha$ and
$\alpha'$, or equivalently, $a$ and $a'$, which we write as
\beq
a = a' f(a') \ ,
\label{aap}
\eeq
where $f(a')$ as the scheme transformation function.  The properties of
the theory must remain unchanged under a scheme transformation in the limit in
which the gauge coupling vanishes and the theory becomes free, which implies
the condition that $f(0) = 1$.  We will use 
a function $f(a')$ that is analytic about $a=a'=0$ and hence has the
power-series expansion 
\beq
f(a') = 1 + \sum_{s=1}^{s_{max}} k_s (a')^s =
        1 + \sum_{s=1}^{s_{max}} \bar k_s (\alpha')^s \ ,
\label{faprime}
\eeq
where the $k_s$ are constants, $\bar k_s = k_s/(4\pi)^s$, and 
$s_{max}$ may be finite or infinite.  The Jacobian of this transformation is 
$J=da/da' = d\alpha/d\alpha'$, with the expansion 
\beq
J = 1 + \sum_{s=1}^{s_{max}} (s+1)k_s(a')^s 
  = 1 + \sum_{s=1}^{s_{max}} (s+1)\bar k_s(\alpha')^s \ . 
\label{j}
\eeq
This Jacobian thus has the value $J=1$ at $a=a'=0$. After the scheme
transformation is applied, the beta function in the resultant scheme is 
\beq
\beta_{\alpha'} \equiv \frac{d\alpha'}{dt} = \frac{d\alpha'}{d\alpha} \,
\frac{d\alpha}{dt} = J^{-1} \, \beta_{\alpha} \ , 
\label{betaap}
\eeq
This has the expansion 
\beq
\beta_{\alpha'} = -2\alpha' \sum_{\ell=1}^\infty b_\ell' (a')^\ell =
-2\alpha' \sum_{\ell=1}^\infty \bar b_\ell' (\alpha')^\ell \ ,
\label{betaprime}
\eeq
with a new set of coefficients $b_\ell'$ (where $\bar b'_\ell =
b'_\ell/(4\pi)^\ell$). One then solves for the $b_\ell'$ as functions of the
$b_\ell$ and $k_s$. This gives $b_1'=b_1$ and $b_2'=b_2$ and the new results
for $b_\ell'$ at higher loop order $\ell$ that were presented in \cite{sch}.
For the reader's convenience, we list some of these results in Appendix
\ref{bellprime_general}.

The $n$-loop beta function in the transformed scheme, $\beta_{\alpha',n\ell}$,
is given by Eq. (\ref{betaprime}) with the upper limit on the $\ell$ summation
equal to $n$ rather than $\infty$. It will be useful to extract the 
quadratic prefactors and define 
\beq
\beta_{\alpha,n\ell,r} \equiv -\frac{\beta_{\alpha,n\ell,r}}
{2\alpha^2} = \sum_{\ell=1}^n \bar b_\ell \, \alpha^{\ell-1} = 
\frac{1}{4\pi} \sum_{\ell=1}^n b_\ell \, a^{\ell-1} 
\label{beta_nloop_reduced}
\eeq
and similarly with $\beta_{\alpha',n\ell,r}$, with the replacements 
$\alpha \to \alpha'$, $b_\ell \to b_\ell'$, and $\bar b_\ell \to \bar b_\ell'$.
Since $b_1'=b_1$ and $b_2'=b_2$, it follows that 
\beq
\beta_{\alpha',2\ell} = \beta_{\alpha,2\ell} \ . 
\label{beta2loopinv}
\eeq
Consequently, if $\beta_{\alpha,2\ell}$ has a (UV or IR) zero at 
$\alpha_{z,2\ell}$, then $\beta_{\alpha',2\ell}$ also has a (UV or IR) zero, 
and at the same value in the transformed variable, 
\beq
\alpha_{z,2\ell}' = \alpha_{z,2\ell} \ . 
\label{aap_2loop}
\eeq
We will use this property below for asymptotically free gauge theories, where
this is an IR zero, so the equality (\ref{aap_2loop}) reads \cite{aapap} 
\beq
\alpha_{IR,2\ell}' = \alpha_{IR,2\ell} = -\frac{4\pi b_1}{b_2} \ . 
\label{alfir_2loop_aap}
\eeq

We recall the set of conditions that a scheme transformation must satisfy in
order to be physically acceptable \cite{sch,sch2}.  The first of these, which
we label as condition $C_1$, is that the scheme transformation must transform a
real positive $\alpha$ to a real positive $\alpha'$, since a function mapping
$\alpha > 0$ to $\alpha'=0$ would be singular, and a function mapping $\alpha >
0$ to a negative or complex $\alpha'$ would violate unitarity. The second
condition, $C_2$, is that the scheme transformation should transform a small or
moderate value of $\alpha$ to a similarly small or moderate value of $\alpha'$,
so a perturbative analysis remains valid.  The third condition, $C_3$, is that
the Jacobian $J$ must be nonzero to avoid a singular transformation
(\ref{betaap}). Since $J=1$ at $\alpha=\alpha'=0$ and $J$ is a continuous
function, condition $C_3$ implies that $J > 0$. The zero of $\beta$ is a
scheme-independent property, and hence, as the fourth condition, $C_4$, a
scheme transformation should be such that $\beta_\alpha$ has a zero if and only
if $\beta_{\alpha'}$ has a zero. The conditions apply for both a scheme
transformation and its inverse. 

These conditions can easily be satisfied by scheme transformations applied in
the vicinity of $\alpha=0$, such as those used to optimize the convergence of
perturbative calculations in quantum chromodynamics \cite{brodskyschemes}, but
they are a significant constraint on a scheme transformation applied in the
vicinity of a (UV or IR) zero of the beta function for $\alpha \lsim
O(1)$. Underlying this analysis of scheme transformations is, of course, the
assumption that one is studying the theory for values of the coupling $\alpha$
that are sufficiently small that perturbative calculations are justified.
Clearly, if the value of $\alpha$ at the zero of the beta function is too
large, then one cannot use perturbative calculational methods reliably. From
the expression for the zero of the beta function, $\alpha_{IR,2\ell}$ in
Eq. (\ref{alfir_2loop}), it is evident that this gets large as $N_f$ decreases
toward the lower end of the interval $I$ at $N_{f,b2z}$ and $b_2$ approaches
zero. Hence, one cannot reliably use perturbative methods to study the
evolution of the coupling near to this lower end of the interval $I$. Since
scheme transformations are carried out in the context of perturbative
calculations, it follows that one could optionally relax the requirement that a
scheme transformation must satisfy all of the conditions $C_1$-$C_4$ at the
lower end of this interval $I$.


\section{General Class of Scheme Transformations $S_{R,m,k_1}$ and 
$S_{R,\infty,k_1}$} 
\label{srmk1}

In this section we present a new scheme transformation $S_{R,m,k_1}$, with $m
\ge 2$ and $s_{max}=m$, that removes the terms in the beta function
$\beta_{\alpha'}$ from loop order $\ell=3$ to order $\ell=m+1$, inclusive. 
In our notation, we have specifically included the value of $k_1$,
since a choice for $k_1$ determines the $k_s$ for $s \ge 2$. Applying the
scheme transformation $S_{R,m,k_1}$ to an initial scheme, it follows that
\beq
S_{R,m,k_1} \ \Longrightarrow \quad  b_\ell'=0 \quad {\rm for} \
\ell=3,...,m+1 \ .
\label{srmk1bell}
\eeq
Thus, $S_{R,m,k_1}$ yields
\beq
\beta_{\alpha',n\ell} = -8\pi(a')^2 \bigg [ b_1 + b_2 a' 
+ \sum_{\ell=m+2}^n b'_\ell (a')^{\ell-1} \bigg ] \ , 
\label{betaprime_nloop_srmk1}
\eeq
and similarly for the expansion in powers of $\alpha$, with $b_\ell'$ replaced
by $\bar b_\ell'$.  From Eq. (\ref{srmk1bell}), it follows that a zero of the
$n$-loop beta function $\beta_{\alpha',m\ell}$ is at the same value as the
(scheme-independent) value $\alpha_{IR,2\ell}$ for $n$ up to and including
$n=m+1$, i.e.,
\beq
S_{R,m} \ \Rightarrow \ \
\alpha_{IR,n\ell}' = \alpha_{IR,2\ell} \quad {\rm for} \ n=3,...,m+1 \ .
\label{alfir_thesamek1}
\eeq

The construction of this scheme makes use of the property that the resultant
coefficient $b'_\ell$ for $\ell \ge 3$ contains only a linear term in
$k_{\ell-1}$, so that the equation $b'_\ell=0$ is a linear equation for
$k_{\ell-1}$, which can always be solved uniquely.  The choice of $k_1$,
together with the values of the $b_\ell$, thus uniquely determines the $k_s$
for $s \ge 2$.  The simplest choice is $k_1=0$, and this was studied in detail
in \cite{sch,sch2}. This special case is indicated with the notation
\beq
S_{R,m,k_1=0} \equiv S_{R,m} \ . 
\eeq

Here we present, as new results, the general formulas for the $k_s$ in the
$S_{R,m,k_1}$ scheme with nonzero $k_1$.  The first step is to use
Eq. (\ref{b3prime}) and solve the equation $b'_3=0$ for $k_2$. This yields
the result
\beq
k_2 = \frac{b_3}{b_1} + \frac{b_2}{b_1} \, k_1 + k_1^2 \quad {\rm for} \ 
S_{R,m,k_1} \ {\rm with} \ m \ge 2 \ . 
\label{k2solk1}
\eeq
This suffices for $S_{R,2,k_1}$. To obtain $S_{R,m,k_1}$ with $m \ge
3$, removing the $\ell=3, \ 4$ terms in $\beta_{\alpha'}$, we need to compute 
$k_3$. For this purpose, we substitute the values of
$k_1$ and $k_2$ into Eq. (\ref{b4prime}) and solve
the equation $b'_4=0$ for $k_3$.  This gives 
\beq
k_3 = \frac{b_4}{2b_1} + \frac{3b_3}{b_1}\, k_1 + \frac{5b_2}{2b_1}\, k_1^2
+ k_1^3 \quad {\rm for} \ S_{R,m,k_1} \ {\rm with} \ m \ge 3 \ . 
\label{k3solk1}
\eeq
Next, to obtain $k_4$, as needed
for $S_{R,m,k_1}$ with $m \ge 4$, we substitute the $k_s$ 
with $s=1, \ 2, \ 3$ into Eq. (\ref{b5prime}) and solve the equation 
$b_5'=0$ for $k_4$. This yields 

\beqs
k_4 & = & \frac{b_5}{3b_1} - \frac{b_2b_4}{6b_1^2} 
   + \frac{5b_3^2}{3b_1^2} + \bigg ( \frac{2b_4}{b_1} + \frac{3b_2b_3}{b_1^2}
\bigg ) k_1 \cr\cr
& + & \bigg ( \frac{6b_3}{b_1} + \frac{3b_2^2}{2b_1^2} \bigg ) k_1^2 
+ \bigg ( \frac{13b_2}{3b_1} \bigg ) k_1^3 + k_1^4  \cr\cr
& & {\rm for} \ S_{R,m,k_1} \ {\rm with} \ m \ge 4 \ .
\label{k4solk1}
\eeqs
We continue this procedure iteratively to calculate $S_{R,m,k_1}$ for higher
$m$.  Thus, having computed the $k_s$ up to order $s=m-1$ inclusive, we compute
$k_m$ by substituting these $k_s$ with $1 \le s \le m-1$ into our expression
for $b_{m+1}'$ and solving the equation $b_{m+1}'=0$ for $k_m$. For a given
$k_1$, this yields a unique solution for $k_m$ because, as noted above, the
equation $b_{m+1}'=0$ with $m+1 \ge 3$ is a linear equation in $k_m$.
Specifically, in the expression for $b_{m+1}'$ with $m+1 \ge 3$, the variable
$k_m$ occurs only in the term $-(m-1)k_m b_1$.  We list the $k_s$ for $s=5$ and
$s=6$ in Appendix \ref{ks_srmk1}. These expressions become progressively
lengthier as $s$ increases, but our method for calculating them as solutions to
respective linear equations is systematic for any $s$.  As is evident, the
choice $k_1=0$ greatly simplifies these expressions for the $k_s$ with $s \ge
2$ and hence also the transformation function $f(a')$.  However, as was shown
in \cite{sch,sch2}, with this choice of $k_1=0$, the scheme transformation
$S_{R,m}$ leads to violations of one or more of the requisite conditions
$C_1$-$C_4$ when applied to the IR zero of the beta function in an
asymptotically free non-Abelian gauge theory with fermions for a substantial
range of $N_f \in I$.  With our generalization, taking advantage of the extra
parameter $k_1$ on which the scheme transformation $S_{R,m,k_1}$ depends, we
obtain a significantly enlarged range of applicability of this scheme
transformation at an IR zero of the beta function.

Because the scheme transformation $S_{R,m,k_1}$ involves coefficients $k_s$
with $s=2,...,m$, the construction of this scheme transformation requires a
knowledge of the $b_\ell$ in this initial scheme up to the loop order
$\ell=m+1$.  Since $s_{max}=m$ for $S_{R,m,k_1}$, it follows that $k_s = 0$ for
$S_{R,m,k_1} $ with $s > m$.  For a given $k_1$, using the $k_s$ with
$s=2,...,m$ as calculated via the procedure above, we compute the $f(a')$
function for the $S_{R,m,k_1}$ scheme transformation:
\beq
f(a')_{S_{R,m,k_1}} = 1 + \sum_{s=1}^m k_s (a')^s 
                    = 1 + \sum_{s=1}^m \bar k_s (\alpha')^s \ . 
\label{faprime_srmk1}
\eeq
Applying this to an initial scheme, we obtain $b_\ell'=0$ for $\ell=3,...,m+1$,
as in (\ref{srmk1bell})-(\ref{betaprime_nloop_srmk1}).

The generalized scheme transformation $S_{R,m,k_1}$ satisfies the same scaling
properties that we derived in \cite{sch} for the case $k_1=0$, i.e., the
$S_{R,m}$ transformation.  Thus, the coefficient $k_s$ depends on the $b_\ell$
with $\ell=1,...,s+1$ via the ratios $b_\ell/b_1$ for $\ell=2,...,s+1$, and
consequently, these $k_s$ are invariant under the rescaling $b_\ell \to \lambda
b_\ell$, where $\lambda \in {\mathbb R}$. It follows that $S_{R,m,k_1}$ is
invariant under the rescaling $b_\ell \to \lambda b_\ell$.  As was true of
$S_{R,m}$, since $S_{R,m,k_1}$ requires knowledge of the $b_\ell$ up to loop
order $\ell=m+1$ and since the $b_\ell$ have been calculated up to $\ell=4$
loops for a general non-Abelian gauge theory \cite{b3,b4}, the highest order
for which we can calculate and apply the $S_{R,m,k_1}$ scheme transformation is
$m=3$.

The application of the transformation $S_{R,m,k_1}$ to an arbitrary initial
scheme yields a $\beta_{\alpha'}$ function with $b_\ell'=0$ for
$\ell=3,...,m+1$, as expressed in
Eqs. (\ref{srmk1bell})-(\ref{betaprime_nloop_srmk1}), so in the new scheme, the
IR zero of the $n$-loop beta function $\beta_{\alpha',m\ell}$ is at the same
value as the (scheme-independent) value $\alpha_{IR,2\ell}$ for $n$ up to and
including $n=m+1$, i.e., $\alpha_{IR,n\ell}' = \alpha_{IR,2\ell}$ for
$n=3,...,m+1$.  

We define $S_{R,\infty,k_1} = \lim_{m \to \infty} S_{R,m,k_1}$. Assuming that
$S_{R,\infty,k_1}$ meets the conditions to be physically acceptable, it takes
an arbitrary initial scheme to a scheme with $b_\ell'=0$ for all $\ell \ge 3$,
so that $\beta_{\alpha'} = -8\pi (a')^2(b_1 + b_2 a') = 
-2(\alpha')^2(\bar b_1 + \bar b_2 \alpha')$. 


\section{Coefficients $b_\ell'$ Resulting from $S_{R,m,k_1}$ Scheme 
Transformation}
\label{bpsrmk1}

\subsection{General Properties}
\label{genprop}

We note some general structural properties of the coefficients $b_\ell'$ for
$S_{R,m,k_1}$.  First, in the expression for $b_\ell'$, the sum of the
subscripts of the $b_\ell$ factors in the numerator of each term minus the
power of $b_1$ in the denominator (if present) plus the power of $k_1$ which
multiplies this term is equal to $\ell$.  For example, in the expression for
the coefficient $b_5'$ resulting from the application of the $S_{R,2,k_1}$
scheme transformation in Eq. (\ref{bp5_sr2k1}) below, in the term
$(12b_2b_3/b_1)k_1$, this sum is $2+3-1+1=5$, and so forth for the other terms
in Eq. (\ref{bp5_sr2k1}) and the other $b_\ell'$.  The (nonzero) coefficient
$b_\ell'$ resulting from the scheme transformation (\ref{faprime}) is, in
general, a polynomial in the $k_s$ for $s=1,...,\ell-1$, and the term in
$b_\ell'$ of highest degree in $k_1$ is proportional to $k_1^{\ell-1}$.  It
follows, in particular, that the term in the nonzero coefficient $b_\ell'$
resulting from the $S_{R,m,k_1}$ scheme transformation (and hence with $\ell
\ge m+2$) is a polynomial in $k_1$ with the property that its highest-degree
term has at most degree $\ell-1$.  Actually, in several cases, the coefficient
of the $k_1^{\ell-1}$ term in $b_\ell'$ vanishes, so the highest-degree term is
proportional to $k_1^{\ell-2}$. This happens, for example, for coefficient
$b_6'$ resulting from the $S_{R,2,k_1}$ scheme transformation and for the
coefficients $b_\ell'$ with $\ell=7, \ 8$ resulting from the $S_{R,3,k_1}$
scheme transformation.


\subsection{$S_{R,2,k_1}$}
\label{bpsr2k1}

Here we give the coefficients $b'_\ell$ resulting from applying the scheme
transformation $S_{R,2,k_1}$ to an initial scheme.  From the expressions for
the $k_s$ in the $S_{R,2,k_1}$ transformation, we obtain the following results
for $s=3, \ 4, \ 5$:
\beq
b'_3=0
\label{bp3_sr2k1} \ , 
\eeq
\beq
b'_4=b_4 + 6b_3k_1 + 5b_2k_1^2 + 2b_1k_1^3 \ , 
\label{bp4_sr2k1}
\eeq
and
\beqs
b'_5 &=& b_5+\frac{5b_3^2}{b_1}+\bigg ( 3b_4+\frac{12b_2b_3}{b_1} \bigg )k_1  
+ \frac{7 b_2^2}{b_1} k_1^2 \cr\cr
& - & b_2k_1^3 - 3b_1k_1^4 \ . 
\label{bp5_sr2k1}
\eeqs
The expressions for $b_\ell'$ for higher $s$ are more lengthy and are given in
Appendix \ref{bellprime_sr2k1}. The expression for the $n$-loop beta function
$\beta_{\alpha',n\ell}$ resulting from the application of the $S_{R,2,k_1}$
transformation is given by the $m=2$ special case of Eq.
(\ref{betaprime_nloop_srmk1}).  


\subsection{$S_{R,3,k_1}$}
\label{bpsr3}

We next present the coefficients $b'_\ell$ resulting from applying the scheme
transformation $S_{R,3,k_1}$ to an initial scheme.  From
the expressions for the $k_s$ in the $S_{R,3,k_1}$ transformation we obtain the
following results for $s=3, \ 4, \ 5$:
\beq 
b'_3 = 0 \ , \quad\quad b'_4 = 0 \ , 
\label{bp34_sr3k1}
\eeq
and
\beqs
b'_5 & = & b_5+\frac{5b_3^2}{b_1}-\frac{b_2b_4}{2b_1} 
+ \bigg ( 6b_4 + \frac{9b_2b_3}{b_1} \bigg ) k_1 \cr\cr
     & + & \bigg ( 18b_3 + \frac{9b_2^2}{2b_1} \bigg ) k_1^2 
     + 13b_2 k_1^3 + 3b_1 k_1^4 \ .
\label{bp5_sr3k1}
\eeqs
We list the expressions for $b_\ell'$ with higher $s$ in Appendix
\ref{bellprime_sr3k1}. The expression for the $n$-loop beta function
$\beta_{\alpha',n\ell}$ following from the application of the $S_{R,3,k_1}$
transformation is given by the $m=3$ special case of Eq.
(\ref{betaprime_nloop_srmk1}).

In a similar manner, one can calculate the coefficients for the $S_{R,m,k_1}$
scheme transformations with $m \ge 4$.  However, to actually apply these scheme
transformations to a given theory requires knowledge of the $b_\ell$
coefficients up to loop order $\ell=m+1$, i.e., $\ell \ge 5$ for $m \ge 4$.
Since our primary application will be to non-Abelian gauge theories, and since
the $b_\ell$ have only been calculated up to loop order $\ell=4$, we thus limit
ourselves to studying the application of the scheme transformations
$S_{R,m,k_1}$ with $m=2$ and $m=3$.


\section{Application of the $S_{R,2,k_1}$ Scheme Transformation}
\label{sr2k1_application}

In this section and the next we discuss the application of the $S_{R,m,k_1}$
scheme transformations.  These transformations can be applied to the beta
function of any gauge theory, non-Abelian or Abelian, asymptotically free or
infrared-free.  As mentioned in the Introduction, we will focus here on the
application to the study of an infrared zero in the beta function of an
asymptotically free vectorial gauge non-Abelian gauge theory with gauge group
$G$ and $N_f$ massless Dirac fermions in a representation $R$ of $G$.  Note
that the two-loop beta function for an Abelian U(1) gauge theory does not have
a zero away from the origin (which would be a UV zero), since $b_1$ and $b_2$
have the same sign (see, e.g., \cite{lnf} and references therein).

In previous work \cite{sch,sch2} it was shown that the special case of the
$S_{R,2,k_1}$ scheme transformation with $k_1=0$, denoted $S_{R,2} \equiv S_2$,
cannot be applied to a generic IR zero of an asymptotically free SU($N_c$)
gauge theory because for a given $N_c$ it fails to satisfy the requisite
conditions to be physically acceptable for a substantial part of the interval
$I$ in Eq. (\ref{nfinterval}).  Here we show that one can pick the parameter
$k_1$ in our generalized one-parameter scheme transformation $S_{R,2,k_1}$ so
as to avoid the pathologies encountered with the $S_{R,2} \equiv 
S_{R,2,k_1=0}$ transformation. 

The $f(a')$ function for the $S_{R,2,k_1}$ scheme transformation is given by
\beqs
S_{R,2,k_1}: \quad f(a') & = & 1 + k_1 a' + \bigg ( \frac{b_3}{b_1} 
+ \frac{b_2}{b_1}k_1 + k_1^2 \bigg ) (a')^2 \cr\cr
& = & 1 + \bar k_1 \alpha' + \bigg ( \frac{\bar b_3}{\bar b_1} 
+ \frac{\bar b_2}{\bar b_1} \bar k_1 + \bar k_1^2 \bigg ) (\alpha')^2 \ , 
\cr\cr
& & 
\label{fap_sr2k1}
\eeqs
and hence the Jacobian is
\beqs
S_{R,2,k_1}: \quad J & = & 1 + 2k_1 a' +
3\bigg ( \frac{b_3}{b_1} + \frac{b_2}{b_1}k_1 + k_1^2 \bigg ) (a')^2 \cr\cr
                     & = & 1 + 2\bar k_1 \alpha' +
 3\bigg ( \frac{\bar b_3}{\bar b_1}
+ \frac{\bar b_2}{\bar b_1} \bar k_1 + \bar k_1^2 \bigg ) (\alpha')^2 \ .
\cr\cr
& &
\label{j_sr2k1}
\eeqs

Now, assume that $N_f \in I$, so that there is an IR zero in the two-loop beta
function, $\beta_{2\ell}$, as given in Eq. (\ref{alfir_2loop}). Since the
existence of an IR zero in beta is a scheme-independent property, one may
impose the condition on an acceptable scheme that it should maintain this
property at higher-loop level.  Because the three-loop expression for the zero
of $\beta_\alpha$ away from the origin involves the square root $\sqrt{b_2^2 -
4b_1b_3}$, and because $b_2 \to 0$ at the smaller-$N_f$ end of the interval
$I$, this condition generically implies that the scheme should be such that
$b_3 < 0$ for $N_f \in I$ \cite{bc}. In particular, this condition is satisfied
in the $\overline{\rm MS}$ scheme \cite{bvh}. We shall impose this condition in
the following.  From our discussion above, it follows that
\beq
\alpha'_{IR,3\ell}=\alpha'_{IR,2\ell}=\alpha_{IR,2\ell} \ , 
\label{alfthesame_sr2k1}
\eeq
provided that the $S_{R,2,k_1}$ transformation is acceptable. 

As in our earlier works \cite{sch,sch2}, the scheme-dependence of the theory in
the vicinity of the IR zero of the beta function is of particular interest, so
we focus on this. The requirement that the $S_{R,2,k_1}$ scheme transformation
should obey condition $C_1$, mapping $a' > 0$ to $a > 0$, is that $f(a') >
0$. This inequality must be satisfied, in particular, at
$a'_{IR,2\ell}=a_{IR,2\ell}=-b_1/b_2$.  Evaluating $f(a')$ at this value, we
obtain
\beq
S_{R,2,k1}: \quad f(a'_{IR,2\ell}) = 1 + \frac{b_1b_3}{b_2^2} 
+ \frac{b_1^2}{b_2^2} k_1^2 \ , 
\label{fap_sr2k1_air2loop}
\eeq
and hence the inequality 
\beq
1 + \frac{b_1b_3}{b_2^2} + \frac{b_1^2}{b_2^2} k_1^2 > 0 \ . 
\label{fap_sr2k1_inequality}
\eeq
(Note that the terms linear in $k_1$ in ( \ref{fap_sr2k1_air2loop}) and
(\ref{fap_sr2k1_inequality}) happen to vanish here and also below in
Eq. (\ref{fxp_sr2k1_xir2loop}).)  Because the coefficient of $k_1^2$ is
positive, this inequality can always be satisfied by using a value of $k_1^2$
that satisfies the inequality
\beq
k_1^2 > (k_1^2)_{min} \ , 
\label{k1sqinequality_for_fap}
\eeq
where 
\beq
(k_1^2)_{min} = -\frac{(b_2^2+b_1b_3)}{b_1^2} = 
\frac{-b_2^2+b_1|b_3|}{b_1^2} \ . 
\label{k1sqmin_fpap}
\eeq
In Eq. (\ref{k1sqmin_fpap}), we have used the property that $b_3 < 0$ for $N_f
\in I$.  By a continuity argument, if $f(a') > 0$ at $a'=a'_{IR,2\ell}$, then
this is also true in a neighborhood of this point on the real $a'$ axis.
Eq. (\ref{k1sqmin_fpap}) is a nontrivial condition if $b_3$ is sufficiently
negative that $|b_3| > b_2^2/b_1$. As was shown in \cite{sch,sch2}, such a
subinterval in $I$ does exist if one uses the $\overline{MS}$ scheme as the
initial scheme. Indeed, this is the reason why $S_{R,2}=S_{R,2,0}$ violates
condition $C_1$.

Condition $C_3$ is that $J > 0$, in particular, at 
$a'_{IR,2\ell}=a_{IR,2\ell}=-b_1/b_2$. Evaluating $J$ at this value, we obtain 
\beq
S_{R,2,k1}: \quad J = 1 + \frac{3b_1b_3}{b_2^2} + \frac{b_1}{b_2}k_1  
+ \frac{3b_1^2}{b_2^2}k_1^2 \ . 
\label{j_sr2k1_air2loop}
\eeq
Then $C_3$ is the inequality 
\beq
1+\frac{3b_1b_3}{b_2^2}+\frac{b_1}{b_2}k_1+\frac{3b_1^2}{b_2^2}k_1^2 > 0 \ .
\label{j_sr2k1_inequality}
\eeq
If $k_1$ were zero, then, since $b_3 < 0$, this condition would be violated for
$|b_3| > b_2^2/(3b_1)$.  For a given $N_c$, as $N_f \in I$ increases and $b_3$
increases in magnitude through negative values, $J$ goes negative before
$f(a')$ does, since $|b_3|$ exceeds $b_2^2/(3b_1)$ before it exceeds
$b_2^2/b_1$. Taking into account that $b_2 < 0$ and $b_3 < 0$ in $I$, the
inequality (\ref{j_sr2k1_inequality}) is satisfied if
\beq
k_1 > \frac{1}{6b_1} \Big ( |b_2| + \sqrt{-11b_2^2 + 36b_1|b_3|} \ \Big )
\label{k1min}
\eeq
or
\beq
k_1 < \frac{1}{6b_1} \Big ( |b_2| - \sqrt{-11b_2^2 + 36b_1|b_3|} \ \Big ) \ .
\label{k1negmax}
\eeq
Note that since we are considering the nontrivial case $|b_3| > b_2^2/(3b_1)$,
the expression in the square root of Eqs. (\ref{k1min}) and (\ref{k1negmax}) is
positive and is greater than $b_1$, which also implies that the right-hand side
of Eq. (\ref{k1negmax}) is negative.  In general, the inequality
(\ref{j_sr2k1_inequality}) is a stronger condition than
(\ref{k1sqinequality_for_fap})-(\ref{k1sqmin_fpap}); for example, with $b_3 <
0$ and $|b_3|=b_2^2/b_1$, it follows that $(k_1^2)_{min}=0$ in
Eq. (\ref{k1sqmin_fpap}), but (\ref{j_sr2k1_inequality}) yields the constraints
that $k_1 > |b_2|/b_1$ from (\ref{k1min}) or $k_1 < -2|b_2|/(3b_1)$ from
(\ref{k1negmax}).

Having shown that $k_1$ can be chosen so that $S_{R,2,k_1}$ satisfies
conditions $C_1$ and $C_3$, we next check conditions $C_3$ and $C_4$.  For this
purpose, we need to analyze the inverse transformation, in which, for a given
$a$, we calculate $a'$ from the relation (\ref{aap}).  For $S_{R,2,k_1}$,
Eq. (\ref{aap}) is the cubic
\beq
S_{R,2,k_1}: \quad a = a'\bigg [ 1 + k_1 a' + \bigg ( \frac{b_3}{b_1} 
+ \frac{b_2}{b_1}k_1 + k_1^2 \bigg ) (a')^2 \bigg ] \ . 
\label{aap_sr2k1}
\eeq

As an illustrative case, we consider $N_c=3$ with $N_f=12$, for which the
two-loop beta function has a (scheme-independent) zero at
$\alpha_{IR,2\ell}=\alpha'_{IR,2\ell}=0.754$, i.e.,
$a_{IR,2\ell}=a'_{IR,2\ell}=0.060$. We study the effect of carrying out the
scheme transformation $S_{R,2,k_1}$ on the beta function. From our general
results above, we calculate $|\bar k_1|_{min} = 0.692$ to satisfy $f(a') > 0$
and $\bar k_1 > 1.525$ or $\bar k_1 < -1.08$ to satisfy $J > 0$. We choose
$\bar k_1 = 1.751$.  Substituting this into Eq. (\ref{aap}) together with
$a=0.060$ and solving for $a'$, we obtain, for the relevant physical root,
$a'=0.0399$, i.e., $\alpha'=0.502$ \cite{aapap}.  (The other two roots of the
cubic are $a'=-0.0575$, which is unphysical, and $a'=0.1107$, which lies
farther away from the origin than $a'=0.0399$ and hence is not reached in the
evolution of the theory from the UV to the IR.)  This moderate shift downward
in the value of the IR zero $\alpha'$ obtained by the $S_{R,2,k_1}$
transformation, is similar to the value of the IR zero that one obtains by
staying within the $\overline{MS}$ scheme and calculating to three loop order,
namely, $\alpha_{IR,3\ell}=0.435$. We have found similar results for other
values of $N_c$ and $N_f$. Thus, condition $C_2$ is satisfied, since the
$S_{R,2,k_1}$ transformation with this value of $k_1$ maps a moderate value of
$a$ to a moderate (smaller) value of $a'$.  Condition $C_4$ is also obviously
satisfied. Continuity of the scheme transformation implies that for values of
$k_1$ close to this value, the same qualitative and quantitative results hold.


\section{Application of the $S_{R,3,k_1}$ Scheme Transformation}
\label{sr3k1_application}

Next, we study the $S_{R,3,k_1}$ scheme transformation. 
The transformation function $f(a')$ for $S_{R,3,k_1}$ is 
\beq
S_{R,3,k_1}: \ f(a') = 1 + k_1 a' + k_2 (a')^2 + k_3 (a')^3 \ , 
\label{fap_sr3k1}
\eeq
where $k_2$ and $k_3$ are given by Eqs. (\ref{k2solk1}) and (\ref{k3solk1}).
From the $m=3$ special case of Eq. (\ref{alfir_thesamek1}), it follows that
after the application of the $S_{R,3}$ scheme transformation, in terms of the
new variable $\alpha'$,
\beq
\alpha'_{IR,4\ell}=\alpha'_{IR,3\ell}=\alpha'_{IR,2\ell}=\alpha_{IR,2\ell} \ .
\label{alfthesame_sr3}
\eeq

We again assume that $N_f \in I$, so that the two-loop beta function has an 
IR zero.  Evaluating $f(a')$ at this (scheme-independent) two-loop zero,
$a'_{IR,2\ell}=a_{IR,2\ell}=-b_1/b_2$, we have
\begin{widetext} 
\beq
S_{R,3,k1}: \quad 
f(a'_{IR,2\ell}) = 1 + \frac{b_1 b_3}{b_2^2} - \frac{b_1^2 b_4}{2b_2^3} 
- 3 \frac{b_1^2b_3}{b_2^3} k_1 - \frac{3b_1^2}{2b_2^2}k_1^2 
- \frac{b_1^3}{b_2^3} k_1^3 \ . 
\label{fap_sr3k1_air2loop}
\eeq
\end{widetext}
An important property of Eq. (\ref{fap_sr3k1_air2loop}) is that the coefficient
of the highest-degree term, $k_1^3$, is positive, namely $-(b_1/b_2)^3 =
(b_1/|b_2|)^3$.  In \cite{sch2} it was shown that for $S_{R,3}=S_{R,3,0}$,
i.e., if $k_1=0$, $f(a'_{IR,2\ell})$ can be negative, violating condition
$C_1$. In contrast, with nonzero $k_1$, because the coefficient of the highest
power of $k_1$ in (\ref{fap_sr3k1_air2loop}) is positive, we can
always satisfy the inequality by using a sufficient large value of $k_1$.

We next consider condition $C_3$, that $J>0$.  Evaluating $J$ at 
$a_{IR,2\ell}' = a_{IR,2\ell}$, we find 
\begin{widetext}
\beq
S_{R,3,k1}: \quad J = 1 + \frac{3b_1b_3}{b_2^2} - \frac{2b_1^2b_4}{b_2^3}
+ \Big ( \frac{b_1}{b_2} - \frac{12b_1^2b_3}{b_2^3} \Big )k_1 
- \frac{7b_1^2}{b_2^2} k_1^2 - \frac{4b_1^3}{b_2^3}k_1^3 \ . 
\label{j_sr3k1_air2loop}
\eeq
\end{widetext}
Again, the coefficient of the highest-degree (degree 3) term in $k_1$, is
positive, namely $-4(b_1/b_2)^3 = 4(b_1/|b_2|)^3$. Hence, we can choose $k_1$
so as to guarantee that $J > 0$ for $N_f \in I$. 

We generalize these results for $S_{R,2,k_1}$ and $S_{R,3,k_1}$ as follows. We
find that for the $S_{R,m,k_1}$ transformation, the respective highest-degree 
terms in the variable $k_1$ in $f(a')$ and $J$ evaluated at $a_{IR,2\ell}'$ 
have degree $m$ and have positive coefficients $\propto (-1)^m (b_1/b_2)^m =
(b_1/|b_2|)^m$. Therefore, by choosing $k_1$ appropriately, one can always
render both $f(a')$ and $J$ evaluated at $a_{IR,2\ell}'$ positive.  This
contrasts with the simpler scheme transformations $S_{R,m} \equiv S_{R,m,0}$
which were analyzed in \cite{sch,sch2} and were shown not to satisfy conditions
$C_1$ and $C_3$.  For values of $a$ that are such that we trust perturbation
theory, the location of the IR zero in $\beta_{n\ell}$ for $n \ge 3$ should not
differ very much from the value in $\beta_{2\ell}$, so by a continuity
argument, it follows that it is possible to choose a $k_1$ that again
guarantees that $f(a')$ and $J$ are positive. In this range of values of $a$,
all of the conditions $C_1$ through $C_4$ are satisfied. 

As noted before, the maximum $m$ for which we can explicitly analyze the
application of the $S_{R,m,k_1}$ scheme transformation in an asymptotically
free theory is $m =3$, because this requires knowledge of the $b_\ell$ for $1
\le \ell \le m+1$, and the $b_\ell$ have only been computed up to $m=4$ loops.
Nevertheless, it is of interest to exhibit the coefficients $b_\ell'$ resulting
from the application of the $S_{R,4,k_1}$ scheme transformation.  We list these
in Appendix \label{bellprime_sr4k1}. 


\section{Scheme Transformations in the Limit 
$N_c \to \infty$, $N_f \to \infty$ with $N_f/N_c$ fixed}
\label{lnn_section}

\subsection{General}

One can get further insight into the application of the $S_{R,2,k_1}$ and
$S_{R,3,k_1}$ scheme transformations at an IR zero of the beta function by
considering an ${\rm SU}(N_c)$ gauge theory with $N_f$ fermions in the
fundamental representation and taking the limit \cite{thooftveneziano} $N_c \to
\infty$ and $N_f \to \infty$ with the ratio
\beq
r \equiv \frac{N_f}{N_c} 
\label{r}
\eeq
held fixed and finite.  One also imposes the condition that the products
\beq
x(\mu) \equiv N_c a(\mu) \ , \quad 
\xi(\mu) \equiv N_c \alpha(\mu) = 4\pi x(\mu) 
\label{xxi}
\eeq
should be fixed, finite functions of $\mu$ in this limit. (As before, we will
often suppress the argument $\mu$ in the notation.) 
We call this the LNN (large $N_c$ and $N_f$) limit.  

As in \cite{lnn}, to have a beta function that has a finite, nontrivial LNN
limit, we multiply both sides of Eq. (\ref{beta}) by $N_c$ and define 
\beq
\beta_{\xi} \equiv \frac{d\xi}{dt} = \lim_{LNN} \beta_\alpha N_c \ .
\label{betaxi}
\eeq
This has the power series expansion
\beq
\beta_\xi \equiv \frac{d\xi}{dt}
= -8\pi x \sum_{\ell=1}^\infty \hat b_\ell x^\ell
= -2 \xi \sum_{\ell=1}^\infty \tilde b_\ell \xi^\ell \ ,
\label{betaxiseries}
\eeq
and 
\beq
  \hat b_\ell = \lim_{LNN} \frac{b_\ell}{N_c^\ell} \ , \quad
\tilde b_\ell = \lim_{LNN} \frac{\bar b_\ell}{N_c^\ell} \ .
\label{bellrel}
\eeq
We define the $n$-loop
$\beta_\xi$ function by Eq. (\ref{betaxiseries}) with the upper limit on the 
summation over loop order $\ell=\infty$ replaced by $\ell=n$.
The (scheme-independent) one-loop and two-loop coefficients in $\beta_\xi$ are
\beq
\hat b_1 = \frac{11-2r}{3} \ , \quad \hat b_2 = \frac{34-13r}{3} \ . 
\label{b12hat}
\eeq
To maintain asymptotic freedom, one restricts $r < 11/2$.  We will focus on the
interval $r \in I_r$ where $\beta_{\xi,2\ell}$ has an IR zero, namely,
\beq
I_r: \quad \frac{34}{13} < r < \frac{11}{2} \ , 
\label{rinterval}
\eeq
i.e., $2.615 < r < 5.500$. This zero occurs at
\beq
x_{IR,2\ell} = \frac{11-2r}{13r-34} \ . 
\label{xir_2loop}
\eeq
We have \cite{lnn}
\beqs
\hat b_3 & = & \frac{1}{54}(2857-1709r+112r^2) \cr\cr
& = & 52.9074-31.6481r+2.07407r^2
\label{b3hat}
\eeqs
and
\begin{widetext}
\beqs
\hat b_4 & = & \frac{150473}{486}-
 \Big ( \frac{485513}{1944} \Big ) r
+\Big ( \frac{8654}{243} \Big ) r^2
+\Big ( \frac{130}{243}  \Big ) r^3 + \frac{4}{9}(11-5r+21r^2) \zeta(3)
\cr\cr
& = & 315.492 - 252.421 \, r + 46.832 \, r^2 + 0.534979 \, r^3 \ ,
\label{b4hat}
\eeqs
\end{widetext}
to the indicated numerical floating-point accuracy, where $\zeta(s) =
\sum_{n=1}^\infty n^{-s}$ is the Riemann $\zeta$ function, 
with $\zeta(3) = 1.202057$.

A scheme transformation in this LNN limit has the form $x = x'f(x')$.  We
impose the condition that $f(0)=1$ to keep the properties of the theory the
same as the coupling goes to zero.  Using an $f(x')$ that is analytic at
$x'=x=0$, we have the expansion
\beq
f(x') = 1 + \sum_{s=1}^{s_{max}} \hat k_s (x')^s 
      = 1 + \sum_{s=1}^{s_{max}} \hat {\bar k_s} (\xi')^s \ . 
\label{fxprime}
\eeq
where the $\hat k_s$ and $\hat {\bar k_s}$ are given by the expressions for the
$k_s$ and $\bar k_s$ with the various $b_n$ coefficients replaced by $\hat
b_n$.  The Jacobian is
\beqs
J & = & \frac{da}{da'} = \frac{dx}{dx'} = 
1 + \sum_{s=1}^{s_{max}} (s+1) \hat k_s (x')^s \cr\cr
  & = & 
1 + \sum_{s=1}^{s_{max}} (s+1) \hat {\bar k_s} (\xi')^s \ . 
\label{jx}
\eeqs
We will denote the scheme transformation on $x$ in the LNN limit that
corresponds to $S_{R,m,k_1}$ with the rescalings indicated above as
$S_{R,m,\hat k_1;LNN}$.  We construct the scheme transformation 
$S_{R,m,\hat k_1;LNN}$ in the same way that we constructed $S_{R,m,k_1}$, 
by solving the equations for $\hat b_\ell=0$ for $3 \le \ell \le m+1$.


\subsection{ $S_{R,2,\hat k_1;LNN}$ Scheme Transformation}
\label{sr2k1_lnn}

For the $S_{R,2,\hat k_1;LNN}$ scheme transformation, we calculate 
\beqs
\hat k_2 & = & \frac{\hat b_3}{\hat b_1} + \frac{\hat b_2}{\hat b_1} \hat k_1
  + \hat k_1^2 \cr\cr
         & = & \frac{2857-1709r+112r^2}{18(11-2r)}
 - \Big (\frac{13r-34}{11-2r}\Big ) \hat k_1 + \hat k_1^2 \ . \cr\cr 
& & 
\label{k2x}
\eeqs
Evaluating the $S_{R,2,\hat k_1;LNN}$ expression for 
$f(x')$ at $x=x_{IR,2\ell}$, we calculate 
\beqs
S_{R,2,\hat k_1;LNN}: & & \ f(x'_{IR,2\ell}) = 
1 + \hat k_1 x'_{IR,2\ell} + \hat k_2 (x'_{IR,2\ell})^2 \cr\cr 
& = & 1 + \frac{\hat b_1 \hat b_3}{\hat b_2^2} + \frac{\hat b_1^2}{\hat b_2^2}
k_1^2 \cr\cr
& = & \frac{52235 - 40425r + 7692r^2 - 224r^3}{18(13r-34)^2} \cr\cr
& + & \bigg ( \frac{11-2r}{13r-34} \bigg )^2 \hat k_1^2 \ . \cr\cr
& & 
\label{fxp_sr2k1_xir2loop}
\eeqs

In \cite{sch2} we showed that for the case $k_1=\hat k_1=0$, i.e., the
$S_{R,2}$ scheme transformation, and $r \in I_r$, $ f(x'_{IR,2\ell})$ is
negative for $34/13 < r < 4.07$ and positive for $4.07 < r < 11/2$ (to the
indicated floating-point numerical accuracy).  Here, by choosing nonzero $\hat
k_1$, we can enlarge the range over which $f(x'_{IR,2\ell}) > 0$, satisfying
condition $C_1$.  The lower bound on $\hat k_1^2$ such that this positivity
holds is
\beq (\hat k_1^2)_{min} = \frac{-52235 + 40425r - 7692r^2 +
224r^3}{18(11-2r)^2} \ .
\label{k1sqmin_sr2k1_ir_lnn}
\eeq
For example, for a value roughly in the middle of the interval $I_r$, namely,
$r=4$, for which $x_{IR,2\ell}=1/6$, this condition is that $|\hat k_1| >
2.12$.

The Jacobian for the $S_{R,2,\hat k_1;LNN}$ scheme transformation, evaluated at
$x'=x'_{IR,2\ell}=-\hat b_1/\hat b_2$, is 
\beqs
S_{R,2,\hat k_1;LNN}: 
J & = & 1 + \frac{3\hat b_1 \hat b_3}{\hat b_2^2} + \frac{\hat b_1}{\hat b_2}
\hat k_1 + \frac{3\hat b_1^2}{\hat b_2^2} \hat k_1^2 \cr\cr
& = & \frac{38363-29817r+5664r^2-224r^3}{6(13r-34)^2} \cr\cr
  & - & \bigg ( \frac{11-2r}{13r-34} \bigg ) \hat k_1 + 
3\bigg ( \frac{11-2r}{13r-34} \bigg )^2 \hat k_1^2 \ . \cr\cr
& & 
\label{j_srk1_ir_lnn}
\eeqs
If $\hat k_1=0$, i.e., for the $S_{R,2}$ scheme transformation, and with $r \in
I_r$, this $J$ is negative for $34/13 < r < 4.69$ and positive for $4.69 < r
< 11/2$.  Here, with the $S_{R,2,k_1}$ scheme transformation, we can choose
$\hat k_1$ to render $J$ positive throughout all of the interval $I_r$, as
required by condition $C_3$.  We can do this because the coefficient of the
term in $J$ of highest degree in $\hat k_1$ (namely, degree 2) is positive.  We
find that $J > 0$ if
\beq
\hat k_1 > \frac{13r-34 + (-75570+58750r-11159r^2+448r^3)^{1/2}}{6(11-2r)} 
\label{k1minj_sr2k1_lnn}
\eeq
or
\beq
\hat k_1 < \frac{13r-34 - (-75570+58750r-11159r^2+448r^3)^{1/2}}{6(11-2r)}
\label{k1maxj_sr2k1_lnn}
\eeq
For example, for a value roughly in the middle of the interval $I_r$, $r=4$,
these inequalities are $\hat k_1 > 6.43$ or $\hat k_1 < -4.43$ (i.e.,
$\hat{\bar k_1} > 0.512$ or $\hat{\bar k_1} < -0.353$).  To check conditions
$C_2$ and $C_4$, we first pick $\hat k_1 = 7$ (i.e., $\hat{\bar k_1}=0.557$)
and substitute this into the equation $x=x'f(x')$ for this $S_{R,2,\hat
k_1;LNN}$ transformation, which is a cubic equation for $x'$. Setting $x$ equal
to the value $x_{IR,2\ell}=1/6$ for $r=4$, and solving for $x'$, we get, as the
relevant physical root, $x'=0.123$.  This is similar to, and slightly smaller
than, $x=1/6=0.167$.  (The other two roots of the cubic equation are
$x'=-0.163$, which is unphysical, and $x'=0.2485$, which is farther from the
origin than $x'=0.123$ and hence is not reached in the evolution of the
coupling from the UV to IR.)  For comparison, we pick $k_1=-6$ and follow the
same procedure.  This yields the relevant physical root $x'=0.179$, slightly
larger than $1/6$.  For both of these choices of $\hat k_1$, all of the
acceptability conditions are satisfied.


\subsection{ $S_{R,3,\hat k_1;LNN}$ Scheme Transformation}
\label{sr3k1_lnn}

The $S_{R,3,\hat k_1;LNN}$ scheme transformation has the same $\hat k_2$ as the
$S_{R,2,\hat k_1;LNN}$ transformation, given above in Eq. (\ref{k2x}).  For
$\hat k_3$, we calculate
\beqs
& & \hat k_3 = \frac{\hat b_4}{2\hat b_1} + \frac{3\hat b_3}{\hat b_1} \hat k_1
  + \frac{5 \hat b_2}{2\hat b_1} \hat k_1^2 + \hat k_1^3 \cr\cr
         & = & \frac{1}{6^4(11-2r)}\bigg [ 601892-485513r+69232r^2+1040r^3
    \cr\cr
& + & \zeta(3)\Big ( 9504-4320r+18144r^2 \Big ) \bigg ] \cr\cr
& + & \frac{(2857-1709r+112r^2)\hat k_1}{6(11-2r)} 
-\frac{5(13r-34)\hat k_1^2}{2(11-2r)} + \hat k_1^3 \ . \cr\cr
& & 
\label{k3x}
\eeqs

The $S_{R,3,\hat k_1;LNN}$ expression for $f(x')$ evaluated at
$x=x_{IR,2\ell}$ is given by the right-hand side of
Eq. (\ref{fap_sr3k1}) with the $b_\ell$ replaced by $\hat b_\ell$ with $1 \le
\ell \le 4$.  Substituting the above expressions for these, we obtain
\begin{widetext}
\beqs
S_{R,3,\hat k_1;LNN} 
\Longrightarrow f(x'_{IR,2\ell}) & = & \frac{1}{6^4(13r-34)^3}
\bigg [ -55042348 + 62622039r - 24520604r^2 + 2885644r^3 + 21504r^4 
+ 4160r^5 \cr\cr
& + & \zeta(3)\Big ( 1149984 - 940896r + 2423520r^2 - 815616r^3 + 72576r^4 
 \Big ) \bigg ] \cr\cr
& + & \frac{(11-2r)^2(2857-1709r+112r^2)\hat k_1}{6(13r-34)^3}
-\frac{3}{2}\bigg ( \frac{11-2r}{13r-34} \bigg )^2 \hat k_1^2 
+ \bigg ( \frac{11-2r}{13r-34} \bigg )^3 \hat k_1^3 \ . 
\label{fxp_sr3k1_lnn}
\eeqs
With the same substitution $x'=x'_{IR,2\ell}$ in $J$, we get 
\beqs
S_{R,3,\hat k_1;LNN} \Longrightarrow J & = & 1 + 
\frac{(11-2r)(2857-1709r+112r^2)}{6(13r-34)^2} \cr\cr
& + & \frac{(11-2r)^2}{324(13r-34)^3}
\bigg [ 601892-485513r+69232r^2+1040r^3 + 
\zeta(3)\Big ( 9504-4320r+18144r^2 \Big ) \bigg ] \cr\cr
& + & \frac{(11-2r)(59386-46374r+8793r^2-448r^3)\hat k_1}{3(13r-34)^3}
-7\bigg ( \frac{11-2r}{13r-34} \bigg )^2 \hat k_1^2 
+4\bigg ( \frac{11-2r}{13r-34} \bigg )^3 \hat k_1^3 \ . 
\cr\cr
& & 
\label{jx_sr3k1_lnn}
\eeqs
\end{widetext}
If $\hat k_1=0$, then for $r \in I_r$, $f(x'_{IR,2\ell})$ is negative for
$34/13 < r < 3.95$ and positive for $3.95 < r < 11/2$, while $J$ is negative
for $34/13 < r < 4.58$ and positive for $4.58 < r < 11/2$.  Since the
coefficients of the $\hat k_1^3$ terms in Eqs. (\ref{fxp_sr3k1_lnn}) and
(\ref{jx_sr3k1_lnn}) are positive, we can choose $\hat k_1$ appropriately to
enlarge the region of $r \in I_r$ for which $f(x_{IR,2\ell})$ and $J$ are
positive, so that conditions $C_1$ and $C_3$ are satisfied. For example, for
the value $r=4$, roughly in the middle of the interval $I_r$,
$f(x'_{IR,2\ell})$ in Eq. (\ref{fxp_sr3k1_lnn}) is positive for $\hat{\bar k_1}
> 1.30$ or $-0.597 < \hat{\bar k_1} < 0.0115$, while $J$ in
Eq. (\ref{jx_sr3k1_lnn}) is positive for $\hat{\bar k_1} > 1.43$ or $-0.543 <
\hat{\bar k_1} < -0.0541$.  Recall that for $r=4$, $x_{IR,2\ell}=1/6$. Setting
$\hat{\bar k_1} = -0.199$ in $f(x')$ for the $S_{R,3,\hat k_1;LNN}$ scheme
transformation and solving the quartic equation $x=x'f(x')$ for this
$S_{R,3,\hat k_1;LNN}$ transformation, we find $x'=0.157$, close to and
slightly smaller than $x_{IR,2\ell}$. (The other three roots of the quartic
equation are all unphysical, namely $x'=-0.190$ and $x'=0.569 \pm 0.142i$.) As
is evident, conditions $C_2$ and $C_4$ are thus also satisfied.  Again one can
use a continuity argument to infer that the same conclusion holds for
neighboring values of $r$ and $\hat k_1$.  Thus, as we did for finite $N_c$ and
$N_f \in I$, here, in the LNN limit with $r \in I_r$, we have shown that, by
the use of the parameter $\hat k_1$ in the $S_{R,2,\hat k_1;LNN}$ and
$S_{R,3,\hat k_1;LNN}$ scheme transformations, we can enlarge the region of
applicability of these transformations as compared with the respective
transformations with $\hat k_1=0$ studied in \cite{sch,sch2}.


\section{On a Modified $S_1$ Scheme Transformation}
\label{s1modified} 

Here we present a modification of the scheme transition denoted $S_1$ in
\cite{sch} which was designed to remove the three-loop term in the beta
function.  This scheme transformation has $s_{max}=1$ and thus has the form
$a=a'(1+k_1 a')$. Solving this quadratic equation for $a'$ formally yields two
solutions, but only one is physical, namely
\beq
a' = \frac{1}{2k_1}\Big ( -1 + \sqrt{1+4k_1a} \ \Big ) \ , 
\label{aap_s1}
\eeq
since only this solution has the property that $a \to a'$ as $a \to 0$.  Since
the purpose of this transformation is to render $b_3'=0$, this condition is
used to determine $k_1$.  The condition $b_3'=0$ in this case is the equation
$b_3 + k_1 b_2 + k_1^2 b_1=0$.  In contrast to the $S_{R,m,k_1}$ scheme
transformation, for which all of the equations for the $k_s$ with $s \ge 2$ are
linear, this equation is quadratic and has the two formal solutions
\beq
k_{1p}, \ k_{1m} = \frac{1}{2b_1}\Big ( -b_2 \pm \sqrt{b_2^2-4b_1b_3} \ \Big )
\label{k1pm}
\eeq
where the $p,m$ subscripts refer to the $\pm$ sign in Eq. (\ref{k1pm}).  If one
requires that this scheme transformation must obey the conditions $C_1$ - $C_4$
throughout all of the interval $I$, then the only acceptable choice is
$k_1=k_{1p}$, as was shown in \cite{sch}.  The application of the $S_1$ scheme
transformation with this choice was analyzed in \cite{sch}.  The regime of
$N_f$ values for which the $S_1$ transformation with $k=k_{1m}$ is unacceptable
is toward the lower end of the interval $I$, where, the value
of the IR zero, $\alpha_{IR,2\ell} = -4\pi b_1/b_2=4\pi b_1/|b_2|$, gets
large. In view of this, one could alternatively choose not to try to apply the
scheme transformation to the lower end of the interval $I$, since one could
plausibly consider that the coupling is too large there for perturbative
methods to be reliable.  In this approach, one could study the application of
the scheme transformation $S_1$ with the choice $k_1 = k_{1m}$ instead of
$k_1=k_{1p}$.

We explore this alternative approach here.  With $b_3 < 0$, we 
reexpress $k_{1m}$ in terms of positive quantities as
\beq
k_{1m} = \frac{1}{2b_1} \Big [ |b_2| - \sqrt{b_2^2+4b_1|b_3|} \ \Big ] \ . 
\label{km1form}
\eeq
If we restricts the application of the $S_1$ scheme transformation to the
middle and upper parts of the interval $I$, then the choice $k_1=k_{1m}$
actually has an advantage as compared with the choice $k_1 = k_{1p}$.  This can
be shown as follows.  We recall that as $N_f$ approaches $N_{f,b1z}$, $b_1$
gets small and consequently, $k_{1p}$ can become somewhat large. This growth in
$k_{1p}$ is cancelled in the $S_1$ transformation, because $k_{1p}$ multiplies
$a'$, and $a$ and $a'$ both approach zero in this limit.  However, this
does lead to some residual scheme dependence in the comparison between
the four-loop IR zero in the $\overline{MS}$ scheme, and the four-loop zero
computed by applying this $S_1$ scheme transformation to that scheme, as
discussed in \cite{sch}. In contrast, with the sign choice $k_1 = k_{1m}$, as 
$N_f$ increases toward $N_{f,b1z}$, $k_{1m}$ approaches $-|b_3|/|b_2|$, and
hence its magnitude does not become large.  Then, taking into account that 
$a_{IR,2\ell}$ approaches zero in this limit, the inversion of the $S_1$ scheme
transformation with $k_1=k_{1m}$ yields values of $a'$ that are closer to the
corresponding values of $a$ in this limit than was the case with the $k_{1p}$
choice. Thus, the $k_{1p}$ and $k_{1m}$ choices have complementary advantages
for the analysis of the IR zero with $N_f \in I$ in these theories. 

\section{Conclusions}
\label{conclusions} 

Because terms at loop order $\ell \ge 3$ in the $\beta$ function of a gauge
theory are scheme-dependent, it follows that one can carry out a scheme
transformation to remove these terms at sufficiently small coupling.  A basic
question concerns the range of applicability of such a scheme transformation.
It is particularly important to address this question when studying the IR zero
that is present in the $\beta$ function of an asymptotically free gauge theory
with sufficiently many fermions.  In this paper we have presented a generalized
class of one-parameter scheme transformations, denoted $S_{R,m,k_1}$ with $m
\ge 2$, depending a parameter $k_1$. A scheme transformation in this class
eliminates the $\ell$-loop terms in the beta function from loop order $\ell=3$
to order $\ell=m+1$, inclusive.  We have analyzed the application of this class
of scheme transformations to the infrared zero of the beta function of a
non-Abelian SU($N_c$) gauge theory with $N_f$ fermions in the fundamental
representation and have shown that an $S_{R,m,k_1}$ scheme transformation in
this class can satisfy the criteria to be physically acceptable over a larger
range of of $N_f$ than the $S_{R,m}$ transformation with $k_1=0$.  As part of
this, we have studied the properties of the corresponding scheme
transformations in the limit $N_c \to \infty$ and $N_f \to \infty$ with
$N_f/N_c$ fixed and finite.  We have also presented and discussed a
modification of the $S_1$ scheme transformation that removes the three-loop
term in the beta of this theory.  These results are useful for the study of the
UV to IR evolution of an asymptotically free gauge theory, and in particular,
the investigation of the properties of a theory of this type with an infrared
fixed point.


\begin{acknowledgments}
This research was partially supported by the NSF grant NSF-PHY-13-16617. 
\end{acknowledgments}


\begin{appendix}

\section{Beta Function Coefficients} 
\label{bell}

For reference, we list the one-loop and two-loop coefficients
\cite{b1,b2,jones} in the beta function (\ref{beta}) for a non-Abelian
vectorial gauge theory with gauge group $G$ and $N_f$ Dirac fermions
transforming according to the representation $R$:
\beq
b_1 = \frac{1}{3}(11 C_A - 4T_fN_f)
\label{b1}
\eeq
\beq
b_2=\frac{1}{3}\left [ 34 C_A^2 - 4(5C_A+3C_f)T_f N_f \right ]
\ .
\label{b2}
\eeq
Our calculations also make use of the three-loop and four-loop coefficients
$b_3$ and $b_4$ calculated \cite{b3,b4} in the $\overline{MS}$ scheme. 


\section{Equations for the $b_\ell'$ Resulting from a General Scheme
Transformation}
\label{bellprime_general}

The expressions for the $b_\ell'$ in Eq. (\ref{betaprime}) 
for $3 \le \ell \le 6$ are \cite{sch} 
\beq
b_3' = b_3 + k_1b_2+(k_1^2-k_2)b_1 
\label{b3prime}
\eeq
\beq
b_4' = b_4 + 2k_1b_3+k_1^2b_2+(-2k_1^3+4k_1k_2-2k_3)b_1 
\label{b4prime}
\eeq
\begin{widetext}
\beq
b_5' = b_5+3k_1b_4+(2k_1^2+k_2)b_3+(-k_1^3+3k_1k_2-k_3)b_2
     + (4k_1^4-11k_1^2k_2+6k_1k_3+4k_2^2-3k_4)b_1 
\label{b5prime}
\eeq
and
\beqs
b_6' & = & b_6 +4k_1b_5+(4k_1^2+2k_2)b_4+4k_1k_2b_3
 + (2k_1^4-6k_1^2k_2+4k_1k_3+3k_2^2-2k_4)b_2 \cr\cr
    & + & (-8k_1^5+28k_1^3k_2-16k_1^2k_3-20k_1k_2^2
      + 8k_1k_4+12k_2k_3 -4k_5)b_1 \ .
\label{b6prime}
\eeqs
The $b_\ell'$ with $\ell$ up to $\ell=8$ were given in \cite{sch}. As was noted
in the text (with $m+1=\ell$), a property that was used in our procedure for
constructing the scheme transformation $S_{R,m,k_1}$ is that in the expressions
for $b_\ell'$ with $\ell \ge 3$, $k_{\ell-1}$ occurs linearly, namely in the
term $-(\ell-2)k_{\ell-1}b_1$.


\section{Higher-Order Coefficients for $S_{R,m,k_1}$} 
\label{ks_srmk1}

In this appendix we list expressions for some higher-order coefficients 
$k_s$ in the $S_{R,m,k_1}$ scheme transformation.  We calculate that 
\beqs
k_5 & = & \frac{b_6}{4b_1} - \frac{b_2b_5}{6b_1^2} + \frac{2b_3b_4}{b_1^2}
+ \frac{b_2^2b_4}{12b_1^3} - \frac{b_2b_3^2}{12b_1^3}
+ \bigg [ \frac{5b_5}{3b_1} + \frac{7b_2b_4}{6b_1^2} + \frac{25b_3^2}{3b_1^2}
  \bigg ] k_1 
+ \bigg [ \frac{5b_4}{b_1} + \frac{27b_2b_3}{2b_1^2} \bigg ] k_1^2 \cr\cr
& + & \bigg [ \frac{10b_3}{b_1} + \frac{35b_2^2}{6b_1^2}  \bigg ] k_1^3 
+ \bigg [\frac{77b_2}{12b_1} \bigg ] k_1^4 + k_1^5 
\quad {\rm for} \ S_{R,m,k_1} \ {\rm with} \ m \ge 5 \ , 
\label{k5solk1}
\eeqs
and
\beqs
k_6 & = & \frac{b_7}{5b_1} - \frac{3b_2b_6}{20b_1^2} + \frac{8b_3b_5}{5b_1^2} 
+ \frac{11b_4^2}{20b_1^2}
- \frac{4b_2b_3b_4}{5b_1^3} + \frac{b_2^2b_5}{10b_1^3} + 
\frac{16b_3^3}{5b_1^3} + \frac{b_2^2b_3^2}{20b_1^4}-\frac{b_2^3b_4}{20b_1^4} 
\cr\cr
& + & \bigg [ \frac{3b_6}{2b_1} + \frac{2b_2b_5}{3b_1^2} + 
\frac{12b_3b_4}{b_1^2} + \frac{47b_2b_3^2}{6b_1^3} - \frac{b_2^2b_4}{3b_1^3} 
\bigg ] k_1 
+ \bigg [ \frac{5b_5}{b_1} + \frac{17b_2b_4}{2b_1^2} + \frac{25b_3^2}{b_1^2} 
+ \frac{15b_2^2b_3}{2b_1^3} \bigg ] k_1^2 \cr\cr
& + & \bigg [ \frac{10b_4}{b_1} + \frac{37b_2b_3}{b_1^2} + 
\frac{5b_2^3}{2b_1^3} \bigg ] k_1^3 + \bigg [ \frac{15b_3}{b_1} + 
\frac{85b_2^2}{6b_1^2} \bigg ] k_1^4 +\bigg [ \frac{87b_2}{10b_1}\bigg ]k_1^5
 + k_1^6
\quad {\rm for} \ S_{R,m,k_1} \ {\rm with} \ m \ge 6 \ .
\label{k6solk1}
\eeqs
%


\section{$b_\ell'$ Coefficients Resulting from the $S_{R,2,k_1}$ Scheme 
Transformation}
\label{bellprime_sr2k1}

From the expressions for $k_s$ in the $S_{R,2,k_1}$ scheme transformation, we
have calculated the resultant coefficients $b'_\ell$ for $\ell$ up to 8.  We
listed $b_\ell'$ for $\ell=3, \ 4, \ 5$ in
Eqs. (\ref{bp3_sr2k1})-(\ref{bp5_sr2k1}) in the text.  Here we give the more
lengthy expressions for the coefficients $b_\ell'$ for $\ell=6, \ 7, \ 8$.  We
have
\beqs
b'_6 & = & b_6+\frac{2b_3b_4}{b_1}+\frac{3b_2b_3^2}{b_1^2} + 
\bigg [ 4b_5 + \frac{2b_2b_4}{b_1} - \frac{16b_3^2}{b_1} + 
\frac{6b_2^2b_3}{b_1^2} \bigg ] k_1 
+ \bigg [ 6b_4 - \frac{36b_2b_3}{b_1} + \frac{3b_2^3}{b_1^2} \bigg ] k_1^2
- \bigg [ 8b_3 + \frac{20b_2^2}{b_1} \bigg ] k_1^3 - 13b_2 k_1^4 \ , \cr\cr
& & 
\label{bp6_sr2k1}
\eeqs
\beqs
b'_7 & = & b_7+\frac{3b_3b_5}{b_1}-\frac{9b_3^3}{b_1^2} 
+ \bigg [ 5b_6 + \frac{3b_2b_5}{b_1} + \frac{7b_3b_4}{b_1} 
- \frac{42b_2b_3^2}{b_1^2} \bigg ] k_1 
+ \bigg [ 10b_5 + \frac{7b_2b_4}{b_1} + \frac{41b_3^2}{b_1} 
- \frac{57b_2^2b_3}{b_1^2} \bigg ] k_1^2 \cr\cr
& + & \bigg [ 9b_4 + \frac{69b_2b_3}{b_1} - \frac{24b_2^3}{b_1^2} \bigg ] k_1^3
+ \bigg [ 44b_3 + \frac{28b_2^2}{b_1} \bigg ] k_1^4 + 41b_2 k_1^5 
+ 9b_1k_1^6\ , 
\label{bp7_sr2k1}
\eeqs
and
\beqs
b'_8 & = & 
b_8+\frac{4b_3b_6}{b_1}+\frac{4b_3^2b_4}{b_1^2}-\frac{8b_2b_3^3}{b_1^3} 
+ \bigg [ 6b_7 + \frac{4b_2b_6}{b_1} + \frac{12b_3b_5}{b_1} 
+ \frac{8b_2b_3b_4}{b_1^2} + \frac{78b_3^3}{b_1^2} 
- \frac{24b_2^2b_3^2}{b_1^3} \bigg ] k_1 \cr\cr
& + & \bigg [ 15b_6+\frac{12b_2b_5}{b_1}+\frac{12b_3b_4}{b_1} 
+ \frac{4b_2^2b_4}{b_1^2} + \frac{258b_2b_3^2}{b_1^2} 
- \frac{24b_2^3b_3}{b_1^3} \bigg ] k_1^2 
+ \bigg [ 18b_5 + \frac{18b_3^2}{b_1}+\frac{12b_2b_4}{b_1}
+\frac{282b_2^2b_3}{b_1^2}-\frac{8b_2^4}{b_1^3} \bigg ] k_1^3 \cr\cr
& + & \bigg [ 9b_4 + \frac{64b_2b_3}{b_1}+\frac{102b_2^3}{b_1^2} \bigg ] k_1^4 
+ \bigg [ -48b_3+\frac{46b_2^2}{b_1} \bigg ] k_1^5 - 42b_2 k_1^6 -18b_1 k_1^7 
\ . 
\label{bp8_sr2k1}
\eeqs
%


\section{$b_\ell'$ Coefficients Resulting from the $S_{R,3,k_1}$ Scheme 
Transformation}
\label{bellprime_sr3k1}

From the expressions for $k_s$ in the $S_{R,3,k_1}$ scheme transformation, we
calculate the resultant $b'_\ell$ coefficients.  We obtain
$b'_3 = 0$, $b'_4 = 0$, and the result for $b'_5$ given in
Eq. (\ref{bp5_sr3k1}).  For the $b'_\ell$ with $\ell=6, \ 7, \ 8$ we find 
\beqs
b'_6 & = & b_6+\frac{8b_3b_4}{b_1}+\frac{3b_2b_3^2}{b_1^2} 
+ \bigg [ 4b_5 + \frac{10b_2b_4}{b_1} + \frac{20b_3^2}{b_1} 
+ \frac{6b_2^2b_3}{b_1^2} \bigg ] k_1
 +  \bigg [ 4b_4 + \frac{42b_2b_3}{b_1} + \frac{3b_2^3}{b_1^2} \bigg ] k_1^2
\cr\cr
& + & \bigg [ -8b_3 + \frac{20b_2^2}{b_1} \bigg ] k_1^3 -7b_2 k_1^4 -4b_1 k_1^5
 \ , 
\label{bp6_sr3k1}
\eeqs
\beqs
b'_7 & = & b_7+\frac{3b_3b_5}{b_1}+\frac{11b_4^2}{4b_1}
-\frac{9b_3^3}{b_1^2}+\frac{9b_2b_3b_4}{2b_1^2}
+ \bigg [ 5b_6 + \frac{3b_2b_5}{b_1} + \frac{10b_3b_4}{b_1} 
- \frac{15b_2b_3^2}{b_1^2} + \frac{9b_2^2b_4}{2b_1^2} \bigg ] k_1 \cr\cr
& + & \bigg [ 10b_5 + \frac{3b_2b_4}{b_1} - \frac{40b_3^2}{b_1} 
-\frac{15b_2^2b_3}{2b_1^2} \bigg ] k_1^2 
+ \bigg [ 10b_4 - \frac{96b_2b_3}{b_1} - \frac{3b_2^3}{2b_1^2} \bigg ] k_1^3
- \bigg [ 10b_3 + \frac{207b_2^2}{4b_1} \bigg ] k_1^4 - 17b_2 k_1^5 \ , 
\label{bp7_sr3k1}
\eeqs
and
\beqs
b'_8 & = & b_8+\frac{4b_3b_6}{b_1}+\frac{b_4b_5}{b_1}-\frac{18b_3^2b_4}{b_1^2}
+\frac{7b_2b_4^2}{4b_1^2}-\frac{8b_2b_3^3}{b_1^3} 
+ \bigg [ 6b_7 + \frac{4b_2b_6}{b_1}+\frac{18b_3b_5}{b_1} 
- \frac{37b_2b_3b_4}{b_1^2} - \frac{54b_3^3}{b_1^2} 
- \frac{24b_2^2b_3^2}{b_1^3} -\frac{15b_4^2}{2b_1} \bigg ] k_1 \cr\cr
& + & \bigg [ 15b_6+\frac{17b_2b_5}{b_1}-\frac{42b_3b_4}{b_1}
- \frac{45b_2^2b_4}{2b_1^2}-\frac{185b_2b_3^2}{b_1^2} 
- \frac{24b_2^3b_3}{b_1^3} \bigg ] k_1^2 
+  \bigg [ 20b_5 - \frac{26b_2b_4}{b_1}-\frac{80b_3^2}{b_1}
- \frac{207b_2^2b_3}{b_1^2} - \frac{8b_2^4}{b_1^3} \bigg ] k_1^3  \cr\cr
&+& \bigg [3b_4-\frac{116b_2b_3}{b_1}-\frac{297b_2^3}{4b_1^2} \bigg ] k_1^4
- \bigg [ 12b_3 + \frac{89b_2^2}{2b_1} \bigg ] k_1^5 - 5b_2 k_1^6 \ . 
\label{bp8_sr3k1}
\eeqs
\end{widetext}

\end{appendix}



\begin{thebibliography}{99}
\bibitem{casimir}
The Casimir invariants $C_R$ and $T_R$ are defined as $\sum_{a=1}^{o(G)}
\sum_{j=1}^{{\rm dim}(R)}{\cal D}_R(T_a)_{ij} {\cal D}_R(T_a)_{jk} = C_R 
\delta_{ik}$ and $\sum_{i,j=1}^{{\rm dim}(R)}
{\cal D}_R(T_a)_{ij} {\cal D}_R(T_b)_{ji} = T_R \delta_{ab}$, where $R$ is the
representation and $T_a$ are the generators of $G$, so that for SU($N_c$),
$C_A=N_c$ for the adjoint ($A$) and $T_{fund}=1/2$ for the fundamental
representation, etc. $C_f$ denotes $C_R$ for the fermion representation.
\bibitem{nfreal}
%
Here and elsewhere, when an analytic expression is given for $N_f$ that 
evaluate to a non-integral real value, it is understood implicitly that 
one infers an appropriate neighboring integral value of $N_f$ from it.
\end{thebibliography}
\end{document}